\documentclass[%
print,
 amsmath,amssymb,
 aps,
]{revtex4}

\usepackage[thmmarks,amsmath]{ntheorem}
\qedsymbol{\ensuremath{\square}}
\newtheorem{theorem}{Theorem}

\newtheorem{example}{Example}

\usepackage[sc]{mathpazo}
\usepackage{graphicx}
\usepackage{dcolumn}
\usepackage{bm}
\usepackage[landscape,papersize={297.1mm,210mm},left=1.9cm,right=1.6cm,top=2.7cm,bottom=2.8cm]{geometry}
\usepackage[                   
            pdfstartview=FitH,
            colorlinks, 
            pdfborder=001,   
            linkcolor=blue,
            anchorcolor=blue,
            citecolor=blue,
            urlcolor=blue
            ]{hyperref}

\usepackage[T1]{fontenc}
\usepackage[english]{babel}
\usepackage{graphicx}
\usepackage{epstopdf}
\usepackage{caption}
\usepackage{subeqnarray}

\usepackage{mathtools}

\begin{document}
\title{Quantifying measurement-induced nonbilocal correlation}

\author{Ying Zhang}
\email{zhangying-1226@163.com}
\affiliation {College of Information and Computer Science, Taiyuan University of Technology, Taiyuan,
 030024, P. R. China}
\author{Kan He}
\email{hekanquantum@163.com}
\affiliation {College of
Mathematics, College of Information and Computer Science, College of
software, Taiyuan University of Technology, Taiyuan,
 030024, P. R. China}

\begin{abstract}
In the paper, we devote to defining an available measure to quantify
the nonbilocal correlation in the entanglement-swapping experiment. Then we obtain analytical formulas to calculate the
quantifier when the inputs are pure states. For the case of mixed
inputs, we discuss the computational properties of the quantifier.
Finally, we derive a tight upper bound to the nonbilocality quantifier.
\end{abstract}


\maketitle

\section{Introduction}
\label{intro}
In the physical interpretation of nature, quantum nonlocality is a fundamental conception and can be applied extensively into quantum information processing for reducing communication complexity, quantum key distribution, private randomness generation, or device-independent entanglement witnesses \cite{CB,MY,ABN,SA,CK,BG}. Nonlocality usually refers to correlations that cannot be described by any local hidden variable theory, and has been widely studied by means of Bell's inequalities focusing on studying correlations between the outcomes of measurements performed on multipartite quantum systems \cite{Bell,4,5,6,7,8}. It is intimately related to, but different from, other strange phenomena such as entanglement and quantumness \cite{9}. It is known that every pure entangled state violates a Bell inequality \cite{Gis0,Pop} and that no separable state does \cite{Werner}, but the situation gets more complicated for mixed entangled states (\cite{Pal,Per,Mas,QZ} and their references). But it is obviously not enough to study nonlocality from the perspective of Bell's inequalites because there is nonlocality without quantumness \cite{35}, just as there is nonlocality without entanglement \cite{36,37,38,39}. Luo and Fu \cite{Luo} define the Measurement-Induced Nonlocality which is in some sense dual to the geometric measure of quantum discord \cite{Dakic} from a geometric perspective in terms of measurements, and obtain analytical formulas for any dimensional pure states and $2\times n$ dimensional mixed states.

Recently, research developments in quantum nonlocality have been inspired by experimental works on quantum networks, where there are multi-measurement and multi-source \cite{BGP,Bra,Fr,Fr2,Wood,Hen,Chaves,Tav,Tav2,Tav3,Chaves2,Rosset}. The simplest example may be the quantum entanglement swapping experiment, in which the joint quantum measurement is the so-called Bell state measurement (BSM). Non-linear inequalities, which allow us to efficiently capture nonbilocal correlations in the entanglement-swapping experiment, were derived \cite{BGP,Bra}. One important class of these inequalities is the so-called binary-input-and-output bilocality inequality which is called the bilocality inequality for simplicity. For pure states, Gisin et al. \cite{Gis} show that every pure entangled state can violate a bilocality inequality, and derive a general criterion which reveals a strong connection between the Clauser-Horne-Shimony-Holt (CHSH) inequality and the bilocality inequality. Particularly, $\rho$ violating CHSH implies $\rho\otimes\rho$ violating the bilocality inequality. In other words, the bilocality inequality can capture more nonlocal correlations.

In this paper, we define a nonbilocality measure in terms of measurements. The paper is structured as follows. In Section \ref{sec:2}, we introduce the definition of the Measure-Induced Nonlocality, and the modified measure. In Section \ref{sec:3}, we define a nonbilocality measure based on measurements and give some basic properties of it. The analytical solutions of the pure states and the tight upper bound of the mixed states are also obtained. In Section \ref{sec:4}, we give three examples to calculate the values of the nonbilocality measure. Finally, we summarize our results in Section \ref{sec:5}.

\section{Relative works}\label{sec:2}
\begin{figure}[!htb]
\centering
\includegraphics[width=3.35in]{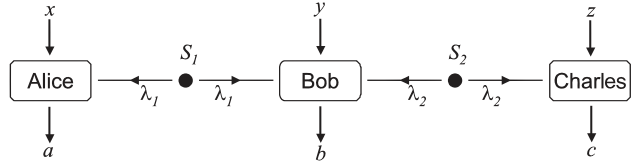}
\caption{Scenario of bilocality}\label{fig2}
\end{figure}

The framework of quantum network corresponding to the bilocal scenario is described in Fig. \ref{fig2}, the $3$-observer and $2$-resource network. In the figure, there are two sources $S_{1}$ and $S_{2}$, where $S_{1}$ distributes two physical systems to distant observers ($A$) and Bob ($B$), and $S_{2}$ distributes two physical systems to Bob ($B$) and Charles ($C$). Consider that Alice receives measurement setting (or input) $x$, while Bob gets input $y$, and Charles $z$. Upon receiving their inputs, each party should provide a measurement result (an output), denoted $a$ for Alice, $b$ for Bob, and $c$ for Charles. The observed statistics is said to be bilocal when
\begin{equation}\label{2.7}
p(a,b,c|x,y,z)=\iint d\lambda_1d\lambda_2 q_1(\lambda_1)q_2(\lambda_2)p(a|x, \lambda_1)p(b|y, \lambda_1, \lambda_2)p(c|z, \lambda_2),
\end{equation}
where $\lambda_1,\lambda_2$ are the independent shared random variables distributed according to the densities $q_1(\lambda_1),q_2(\lambda_2)$.

If we consider that Alice and Charles receive binary inputs, $x=0,1$ and $z=0,1$, and must give binary outputs, denoted $A_x=\pm1$ and $C_z=\pm1$, respectively. The middle party Bob always performs the same measurement (no input $y$) with four possible outcomes, as, e.g., the Bell state measurement. Denote Bob's outcome by two bits $B_0=\pm1$ and $B_1=\pm1$. Then, the bilocality inequality is
obtained from Eq. \eqref{2.7} as follows
\begin{equation}\label{2.8}
S\equiv\sqrt{|I|}+\sqrt{|J|}\leq 2,
\end{equation}
where
\begin{equation*}
\begin{aligned}
I\equiv\langle(A_0+A_1)B_0(C_0+C_1)\rangle,\\
J\equiv\langle(A_0-A_1)B_0(C_0-C_1)\rangle.
\end{aligned}
\end{equation*}
The bracket $\langle\cdot\rangle$ denotes the expectation value of many experimental runs.

Consider a bipartite system with composite Hilbert space $H=H_{A}\otimes H_{B}$. Let $\mathcal{D}(H)$ be the set of bounded, positive-semidefinite operators with unit trace on $H$. Given a quantum state $\rho_{AB}\in \mathcal{D}(H)$ which could be shared between two parties, Alice and Bob, let $\rho_{A}$ and $\rho_{B}$ be the reduced density matrix for each party. For the bipartite $\rho_{AB}$,
 Alice performs local von Neumann measurements which do not disturb the local state $\rho_{A}=\mathrm{tr}_{B}\rho_{AB}$. To capture all the nonlocal effects that can be induced by local measurements, Luo and Fu define the Measurement-Induced Nonlocality \cite{Luo} in terms of the Hilbert-Schmidt norm,
\begin{equation}\label{3.1}
\begin{aligned}
N(\rho_{AB})\equiv\max_{\Pi^{A}} \|\rho_{AB}-\Pi^{A}(\rho_{AB})\|^{2},
\end{aligned}
\end{equation}
where the maximum is taken over all the von Neumann measurements $\{\Pi^{A}\}$ which do not disturb $\rho_{A}$ locally, that is, $\rho_{A}=\sum_{k}\Pi^{A}_{k}\rho_{A}\Pi^{A}_{k}$ and $\Pi^{A}(\rho_{AB})=\sum_{k}(\Pi^{A}_{k}\otimes I^{B})\rho_{AB}(\Pi^{A}_{k}\otimes I^{B})$. The post-measurement state $\Pi^{A}(\rho_{AB})$ can be rewritten as
$$\Pi^{A}(\rho_{AB})=\sum_{k}p_{k}\Pi_{k}^{A}\otimes\rho_{k}^{B},$$
where $\rho_{k}^{B}=\frac{1}{p_{k}}\mathrm{tr}_{A}(\Pi^{A}_{k}\otimes I^{B}\rho_{AB})$ is the post-measurement state of system $B$ that corresponds to the probability $p_{k}=\mathrm{tr}(\Pi^{A}_{k}\otimes I^{B}\rho_{AB})$. Here, the Hilbert-Schmidt norm is defined as $\|X\|\equiv\sqrt{\mathrm{tr}(X^{\dag}X)}$.

This measure is in some sense dual to, the geometric measure of quantum discord \cite{Dakic}
\begin{equation}
\begin{aligned}
D(\rho_{AB})\equiv\min_{\Pi^{A}} \|\rho_{AB}-\Pi^{A}(\rho_{AB})\|^{2},
\end{aligned}
\end{equation}
where the minimum is taken over all the von Neumann measurements $\{\Pi^{A}\}$ which do not disturb $\rho_{A}$ locally.

However, this measure has a curious drawback in that it may change reversibly by trivially adding a local ancilla \cite{41}. Chang and Luo \cite{42} modified this measure as follows
\begin{equation}
\begin{aligned}
D_H(\rho_{AB})\equiv\min_{\Pi^{A}} \|\sqrt{\rho_{AB}}-\Pi^{A}(\sqrt{\rho_{AB}})\|^{2},
\end{aligned}
\end{equation}
where $\Pi^A(\sqrt{\rho_{AB}})\equiv\sum_k(\Pi_k^A\otimes I^B)\sqrt{\rho_{AB}}(\Pi_k^A\otimes I^B)$.

So it is necessary to discuss the modified Measurement-Induced Nonlocality where the $\rho_{AB}$ is replaced by its square root $\sqrt{\rho_{AB}}$. In the following, we study the nonbilocal correlation by giving a nonbilocality measure from the perspective of measurements, and give a connection between the modified Measurement-Induced Nonlocality and the nonbilocality measure.
\section{Quantifying the nonbilocal correlation}\label{sec:3}

For an arbitrary finite dimensional system $H=H_A\otimes H_B\otimes
H_C\otimes H_D$, define the nonbilocality measure as
\begin{equation}\label{4.1}
N_H^{b}(\rho_{AB}\otimes\rho_{CD})= \max_{\Pi^{BC}} \|\sqrt{\rho_{AB}\otimes\rho_{CD}}-\Pi^{BC}(\sqrt{\rho_{AB}\otimes \rho_{CD}})\|^2,
\end{equation}
where the max is taken over all the von Neumann measurements $\Pi^{BC}=\{\Pi_h^{BC}\}$ which do not disturb $\rho_{BC}$ locally, that is, $\sum_h \Pi_h^{BC}\rho_{BC}\Pi_h^{BC} =\rho_{BC}$, and $\|\cdot\|$ is the Hilbert-Schmidt norm, $\Pi^{BC}(\sqrt{\rho_{AB}\otimes \rho_{CD}})\equiv\sum_h(I^{A}\otimes\Pi_h^{BC}\otimes I^D)\sqrt{\rho_{AB}\otimes \rho_{CD}}(I^{A}\otimes\Pi_h^{BC}\otimes I^D)$. Let $\rho_{A}$ and $\rho_{B}$ be the reduced density matrices of $\rho_{AB}$ for party $A$ and party $B$ respectively, and the same for $\rho_{C},\rho_{D}$. Then we have $\rho_{BC}={\rm tr}_{AD}(\rho_{AB}\otimes\rho_{CD})=\rho_{B}\otimes\rho_{C}.$
\subsection{Basic properties}
Now we list some basic properties of the nonbilocality measure.

(1) $N_H^{b}(\rho_{AB}\otimes\rho_{CD})=0$ for any product states $\rho_{AB}=\rho_{A}\otimes\rho_{B}$ and $\rho_{CD}=\rho_{C}\otimes\rho_{D}$. This follows from the fact that if $\sum_h\Pi_h^{BC}\rho_{BC}\Pi_h^{BC}=\rho_{BC}$, then $\rho_{BC}$ must have a spectral decomposition $\rho_{BC}=\sum_h p_h\Pi_h^{BC}$. Consequently, $\sqrt{\rho_{BC}}=\sum_h\sqrt{p_h}\Pi_h^{BC}=\sum_h\Pi_h^{BC}\sqrt{\rho_{BC}}\Pi_h^{BC}$.

(2) $N_H^{b}(\rho_{AB}\otimes\rho_{CD})$ is locally unitary invariant in the sense that $N_H^{b}((U_{A}\otimes U_{B}\otimes U_{C}\otimes U_{D})(\rho_{AB}\otimes\rho_{CD})(U_{A}\otimes U_{B}\otimes U_{C}\otimes U_{D})^{\dag})=N_H^{b}(\rho_{AB}\otimes\rho_{CD})$ for any unitary operators $U_{A}, U_{B}, U_{C}$ and $U_{D}$ acting on $H_A,H_B,H_C$ and $H_D$, respectively. This follows readily from the fact that \begin{equation*}
\begin{aligned}
&\sqrt{(U_{A}\otimes U_{B}\otimes U_{C}\otimes U_{D})(\rho_{AB}\otimes\rho_{CD})(U_{A}\otimes U_{B}\otimes U_{C}\otimes U_{D})^{\dag}}=\\
&(U_{A}\otimes U_{B}\otimes U_{C}\otimes U_{D})\sqrt{\rho_{AB}\otimes\rho_{CD}}(U_{A}\otimes U_{B}\otimes U_{C}\otimes U_{D})^{\dag}.
\end{aligned}
\end{equation*}

(3) If $\rho_{B}$, $\rho_{C}$ have the spectral decomposition $\rho_{B}=\sum_{e}\lambda_{e}|e_{B}\rangle\langle e_{B}|$, $\rho_{C}=\sum_{f}\mu_{f}|f_{C}\rangle\langle f_{C}|$ respectively and at least one of the states $\rho_{B}$ and $\rho_{C}$ is nondegenerate, then the von Neumann measurements that do not disturb $\rho_{BC}=\rho_{B}\otimes\rho_{C}$ cannot be the joint quantum measurements, and they must have the form $\Pi^{BC}=\{\Pi_e^{B}\otimes\Pi_f^{C}\}$.

(4) If $\rho_{B}$, $\rho_{C}$ are both nondegenerate with spectral decomposition $\rho_{B}=\sum_{e}\lambda_{e}|e_{B}\rangle\langle e_{B}|$ and $\rho_{C}=\sum_{f}\mu_{f}|f_{C}\rangle\langle f_{C}|$ respectively, then $N_H^{b}(\rho_{AB}\otimes\rho_{CD})=\|\sqrt{\rho_{AB}\otimes\rho_{CD}}-\Pi^{BC}(\sqrt{\rho_{AB}\otimes \rho_{CD}})\|^2$ with $\Pi^{BC}(\rho_{AB}\otimes \rho_{CD})=\sum_{ef}(I^{A}\otimes\Pi_{e}^{B}\otimes\Pi_{f}^{C}\otimes I^{D})(\rho_{AB}\otimes \rho_{CD})(I^{A}\otimes\Pi_{e}^{B}\otimes\Pi_{f}^{C}\otimes I^{D})=\sum_{ef}(I^{A}\otimes|e_{B}\rangle\langle e_{B}|\otimes|f_{C}\rangle\langle f_{C}|\otimes I^{D})(\rho_{AB}\otimes \rho_{CD})(I^{A}\otimes|e_{B}\rangle\langle e_{B}|\otimes|f_{C}\rangle\langle f_{C}|\otimes I^{D})$. This is because in such a situation, the only von Neumann measurement that does not disturb $\rho_{BC}=\rho_{B}\otimes\rho_{C}$ is $\Pi^{BC}=\{\Pi_e^{B}\otimes\Pi_f^{C}=|e_{B}\rangle\langle e_{B}|\otimes|f_{C}\rangle\langle f_{C}|\}$, and thus the max in Eq. \eqref{4.1} is not necessary. In particular, $N_{H}^{b}(\rho_{AB}\otimes\rho_{CD})$ vanishes for any classical-quantum states $\rho_{AB}=\sum_{e}\rho_{e}^{A}\otimes p_{e}|e_{B}\rangle\langle e_{B}|$ and $\rho_{CD}=\sum_{f} q_{f}|f_{C}\rangle\langle f_{C}|\otimes \rho_{f}^{D}$ whose marginal states $\rho_{B}=\sum_{e}p_{e}|e_{B}\rangle\langle e_{B}|$ and $\rho_{C}=\sum_{f}q_{f}|f_{C}\rangle\langle f_{C}|$ are both nondegenerate.

(5) $N_H^{b}(\rho_{AB}\otimes\rho_{CD})$ is strictly positive for any input states $\rho_{AB},\rho_{CD}$ where at least one of the states is entanglement. Suppose $N_H^{b}(\rho_{AB}\otimes\rho_{CD})=0$, then for some von Neumann measurement $\Pi^{BC}$, we have the equation $\sqrt{\rho_{AB}\otimes\rho_{CD}}=\Pi^{BC}(\sqrt{\rho_{AB}\otimes \rho_{CD}})$ holds. Noting that $\Pi^{BC}(\sqrt{\rho_{BC,AD}})$ must have the form $\Pi^{BC}(\sqrt{\rho_{BC,AD}})=\sum_h\Pi_h^{BC}\otimes \sqrt{p_h\rho_h^{AD}}=\sqrt{\rho_{BC,AD}}$. That implies that $\rho_{BC,AD}=\sum_h\Pi_h^{BC}\otimes p_h\rho_h^{AD}=\Pi^{BC}(\rho_{BC,AD})$. Thus the states $\rho_{AB}$ and $\rho_{CD}$ are both separable which leads to a contradiction.

(6) We get a strong connection between modified Measure-Induced nonlocality and our measure, that is, $N_{H}(\rho_{AB})>0$ implies $N_{H}^{b}(\rho_{BA}\otimes\rho_{AB})>0$ since

\begin{equation*}
\begin{aligned}
&N_{H}^{b}(\rho_{BA}\otimes\rho_{AB})\\
=&\max_{\Pi^{AA}}\|\sqrt{\rho_{BA}\otimes\rho_{AB}}-\Pi^{AA}(\sqrt{\rho_{BA}\otimes\rho_{AB}})\|^{2}\\
\geq&\max_{\Pi^{A}}\|\sqrt{\rho_{BA}\otimes\rho_{AB}}-(\Pi^{A}\otimes\Pi^{A})(\sqrt{\rho_{BA}\otimes\rho_{AB}})\|^2\\
=&\max_{\Pi^{A}}\|\sqrt{\rho_{BA}}\otimes\sqrt{\rho_{AB}}-\Pi^{A}(\sqrt{\rho_{BA}})\otimes\Pi^{A}(\sqrt{\rho_{AB}})\|^{2}\\
=&(\mathrm{tr}(\rho_{AB}))^2-\min_{\Pi^{A}}\mathrm{tr}(\sqrt{\rho_{BA}}\otimes\sqrt{\rho_{AB}})(\Pi^{A}(\sqrt{\rho_{BA}})\otimes\Pi^{A}(\sqrt{\rho_{AB}}))\\
=&1-\min_{\Pi^{A}}(\mathrm{tr}\sqrt{\rho_{AB}}\Pi^{A}(\sqrt{\rho_{AB}}))^{2}\\
\geq&1-\min_{\Pi^{A}}~\mathrm{tr}\sqrt{\rho_{AB}}\Pi^{A}(\sqrt{\rho_{AB}})\\
=&N_{H}(\rho_{AB}).
\end{aligned}
\end{equation*}
\subsection{The case of pure states}
The nonbilocality measure for any pure states can be evaluated as follows.

\begin{theorem}\label{theorem1}
If $\rho_{AB}\otimes
\rho_{CD}=|\psi_{AB}\rangle\langle \psi_{AB}|\otimes|
\phi_{CD}\rangle\langle \phi_{CD}|$ is pure, then
$$N_H^{b}(\rho_{AB}\otimes
\rho_{CD})=1-\sum_{ij} \lambda_i^4 \mu_j^4,$$ where $\lambda_i$s
and $\mu_j$s are Schmidt coefficients of $|\psi_{AB}\rangle$ and $|\phi_{CD}\rangle$, respectively.
\end{theorem}

\begin{proof*}
Noting that $|\psi_{AB}\rangle=\sum_i\lambda_i|i_Ai_B\rangle,|\phi_{CD}\rangle=\sum_j\mu_j|j_Cj_D\rangle,$
\begin{equation*}
\begin{aligned}
\rho_{AB}\otimes \rho_{CD}&=|\psi_{AB}\rangle\langle\psi_{AB}|\otimes|\phi_{CD}\rangle\langle\phi_{CD}|\\
&=\sum_{ii^{'}jj^{'}}\lambda_i\lambda_{i^{'}}\mu_j\mu_{j^{'}}|i_A\rangle\langle i^{'}_A|\otimes |i_B\rangle\langle i^{'}_B|\otimes |j_C\rangle\langle j^{'}_C|\otimes |j_D\rangle\langle j^{'}_D|,
\end{aligned}
\end{equation*}
and
\begin{equation*}
\begin{aligned}
\rho_{BC}&={\rm tr}_{AD}(|\psi_{AB}\rangle\langle\psi_{AB}|\otimes|\phi_{CD}\rangle\langle\phi_{CD}|)=\sum_{ij}\lambda_i^2\mu_j^2|i_Bj_C\rangle\langle i_Bj_C|,
\end{aligned}
\end{equation*}
$\Pi^{BC}(\sqrt{\rho_{AB}\otimes \rho_{CD}})$ can be rewritten as

\begin{equation*}
\begin{aligned}
&\Pi^{BC}(\sqrt{\rho_{AB}\otimes \rho_{CD}})=\Pi^{BC}(\rho_{AB}\otimes \rho_{CD})\\
=&\sum_{ef}(I^{A}\otimes\Pi_{ef}^{BC}\otimes
I^{D})(|\psi_{AB}\rangle\langle\psi_{AB}|\otimes|\phi_{CD}\rangle\langle\phi_{CD}|)(I^{A}\otimes\Pi_{ef}^{BC}\otimes I^{D})\\
=&\sum_{ef}(I^{A}\otimes\Pi_{ef}^{BC}\otimes I^{D})(\sum_{ii^{'}jj^{'}}\lambda_i\lambda_{i^{'}}\mu_j\mu_{j^{'}}|i_A\rangle\langle i^{'}_A|\otimes |i_B\rangle\langle i^{'}_B|\otimes |j_C\rangle\langle j^{'}_C|\otimes|j_D\rangle\langle j^{'}_D|)(I^{A}\otimes\Pi_{ef}^{BC}\otimes I^{D})\\
=&\sum_{ef}\sum_{ii^{'}jj^{'}}\lambda_i\lambda_{i^{'}}\mu_j\mu_{j^{'}}|i_A\rangle\langle i^{'}_A|\otimes \Pi^{BC}_{ef}|i_B j_{C}\rangle\langle i^{'}_B j^{'}_C|\Pi^{BC}_{ef}\otimes |j_D\rangle\langle j^{'}_D|\\
=&\sum_{ef}\sum_{ii^{'}jj^{'}}\lambda_i\lambda_{i^{'}}\mu_j\mu_{j^{'}}|i_A\rangle\langle i^{'}_A|\otimes U|e_{B}f_{C}\rangle\langle e_{B}f_{C}|U^{\dag}|i_B j_{C}\rangle\langle i^{'}_B j^{'}_C|U|e_{B}f_{C}\rangle\langle e_{B}f_{C}|U^{\dag}\otimes |j_D\rangle\langle j^{'}_D|\\
=&\sum_{ef}\sum_{ii^{'}jj^{'}}\lambda_i\lambda_{i^{'}}\mu_j\mu_{j^{'}}\langle e_{B}f_{C}|U^{\dag}|i_B j_{C}\rangle\langle i^{'}_B j^{'}_C|U|e_{B}f_{C}\rangle|i_A\rangle\langle i^{'}_A|\otimes U|e_{B}f_{C}\rangle\langle e_{B}f_{C}|U^{\dag}\otimes |j_D\rangle\langle j^{'}_D|,
\end{aligned}
\end{equation*}
where any von Neumann measurement on $H_{B}\otimes H_{C}$ is expressed as $\Pi^{BC}=\{\Pi^{BC}_{ef}\equiv U|e_{B}f_{C}\rangle\langle e_{B}f_{C}|U^{\dag}\}$. Consequently,
\begin{equation*}
\begin{aligned}
&\sqrt{\rho_{AB}\otimes \rho_{CD}}\Pi^{BC}(\sqrt{\rho_{AB}\otimes \rho_{CD}})\\
=&(\sum_{ii^{'}jj^{'}}\lambda_i\lambda_{i^{'}}\mu_j\mu_{j^{'}}|i_A\rangle\langle i^{'}_A|\otimes |i_B\rangle\langle i^{'}_B|\otimes |j_C\rangle\langle j^{'}_C|\otimes |j_D\rangle\langle j^{'}_D|)(\sum_{ef}\sum_{uu^{'}vv^{'}}\lambda_u\lambda_{u^{'}}\\
&\mu_v\mu_{v^{'}}\langle e_{B}f_{C}|U^{\dag}|u_B v_{C}\rangle\langle u^{'}_B v^{'}_C|U|e_{B}f_{C}\rangle|u_A\rangle\langle u^{'}_A|\otimes U|e_{B}f_{C}\rangle\langle e_{B}f_{C}|U^{\dag}\otimes \\
&|v_D\rangle\langle v^{'}_D|)\\
=&\sum_{ii^{'}jj^{'}}\sum_{ef}\sum_{uu^{'}vv^{'}}\lambda_i\lambda_{i^{'}}\mu_j\mu_{j^{'}}\lambda_u\lambda_{u^{'}}\mu_v\mu_{v^{'}}\langle e_{B}f_{C}|U^{\dag}|u_B v_{C}\rangle\langle u^{'}_B v^{'}_C|U|e_{B}f_{C}\rangle|i_A\rangle\\
&\langle i^{'}_A|u_A\rangle\langle u^{'}_A|\otimes|i_B j_C\rangle\langle i^{'}_B j^{'}_C|U|e_{B}f_{C}\rangle\langle e_{B}f_{C}|U^{\dag}\otimes |j_D\rangle\langle j^{'}_D|v_D\rangle\langle v^{'}_D|,
\end{aligned}
\end{equation*}
from which we obtain
\begin{equation*}
\begin{aligned}
&\mathrm{tr}\sqrt{\rho_{AB}\otimes \rho_{CD}}\Pi^{BC}(\sqrt{\rho_{AB}\otimes \rho_{CD}})\\
=&\sum_{iujvef}\lambda_i^2\lambda_{u}^2\mu_j^2\mu_{v}^2\langle e_{B}f_{C}|U^{\dag}|u_B v_{C}\rangle\langle i_B j_C|U|e_{B}f_{C}\rangle\langle u_B v_C|U|e_{B}f_{C}\rangle\langle e_{B}f_{C}|U^{\dag}|i_B j_C\rangle\\
=&\sum_{ef}(\sum_{ij}\lambda_i^2\mu_j^2\langle i_B j_C|U|e_{B}f_{C}\rangle\langle e_{B}f_{C}|U^{\dag}|i_B j_C\rangle)^{2}\\
=&\sum_{ef}(\langle e_{B}f_{C}|U^{\dag}\rho_{BC} U|e_{B}f_{C}\rangle)^{2}.
\end{aligned}
\end{equation*}
Now
\begin{equation*}
\begin{aligned}
N_H^{b}(\rho_{AB}\otimes
\rho_{CD})=& \max_{\Pi^{BC}} \|\sqrt{\rho_{AB}\otimes
\rho_{CD}}-\Pi^{BC}(\sqrt{\rho_{AB}\otimes \rho_{CD}})\|^2\\
=& \max_{\Pi^{BC}} \|\rho_{AB}\otimes
\rho_{CD}-\Pi^{BC}(\rho_{AB}\otimes \rho_{CD})\|^2\\
=&\max_{\Pi^{BC}} (\mathrm{tr}(\rho_{AB}\otimes \rho_{CD})^{2}-\mathrm{tr}(\rho_{AB}\otimes \rho_{CD})\Pi^{BC}(\rho_{AB}\otimes \rho_{CD}))\\
=&1-\min_{\Pi^{BC}}\sum_{ef}(\langle e_{B}f_{C}|U^{\dag}\rho_{BC} U|e_{B}f_{C}\rangle)^{2},
\end{aligned}
\end{equation*}
where the optimization is over all von Neumann measurements $\Pi^{BC}=\{U|e_{B}f_{C}\rangle$
$\langle e_{B}f_{C}|U^{\dag}\}$ leaving the marginal state $\rho_{BC}$ invariant. This invariance means that
$$\sum_{ef}U|e_{B}f_{C}\rangle\langle e_{B}f_{C}|U^{\dag}\rho^{BC}U|e_{B}f_{C}\rangle\langle e_{B}f_{C}|U^{\dag}=\rho_{BC},$$
or equivalently,
$$\rho_{BC}=\sum_{ef}\langle e_{B}f_{C}|U^{\dag}\rho_{BC}U|e_{B}f_{C}\rangle U|e_{B}f_{C}\rangle\langle e_{B}f_{C}|U^{\dag}$$
is a spectral decomposition of $\rho_{BC}$ since $\{U|e_{B}f_{C}\rangle\}$ is an orthonormal base. Consequently,
$$\sum_{ef}(\langle e_{B}f_{C}|U^{\dag}\rho_{BC} U|e_{B}f_{C}\rangle)^{2}=\sum_{ef} \lambda_e^4 \mu_f^4,$$
and the desired result is obtained. The optimum is actually achieved by any von Neumann measurement leaving $\rho_{BC}$ invariant.
\qed
\end{proof*}

\subsection{The case of mixed states}
How about mixed states? Suppose the Hilbert spaces $H_A,H_B,H_C,$ and $H_D$ are of dimensions $\dim H_A=m,\dim H_B=n,\dim H_C=u,$ and $\dim H_C=v$, respectively. Let $L(H^{x})$ be the Hilbert space consisting of all linear operators on $H^{x}(x=A,B,C,D)$, with the Hilbert-Schmidt inner product $\langle X|Y\rangle\equiv\text{tr}X^\dag Y$. Let $\{X_{i}:i=0,1,\ldots,m^{2}-1\}$, $\{Y_{j}:i=0,1,\ldots,n^{2}-1\}$, $\{Z_{k}:k=0,1,\ldots,u^{2}-1\}$ and $\{W_{l}:l=0,1,\ldots,v^{2}-1\}$ be orthonormal Hermitian operator bases for $L(H_{A})$, $L(H_{B})$, $L(H_{C})$ and $L(H_{D})$ respectively, with $X_{0}=I^{A}/\sqrt{m}$, $Y_{0}=I^{B}/\sqrt{n}$, $Z_{0}=I^{C}/\sqrt{u}$ and $W_{0}=I^{D}/\sqrt{v}$. Then general bipartite states $\rho_{AB}$ and $\rho_{CD}$ can always be represented as
\begin{equation}\label{4.2}
\begin{aligned}
\sqrt{\rho_{AB}}=\sum_{i=0}^{m^{2}-1}\sum_{j=0}^{n^{2}-1}\gamma_{ij}^{AB}X_{i}\otimes Y_{j}, \quad \sqrt{\rho_{CD}}=\sum_{k=0}^{u^{2}-1}\sum_{l=0}^{v^{2}-1}\gamma_{kl}^{CD}Z_{k}\otimes W_{l},
\end{aligned}
\end{equation}
where $\gamma_{ij}^{AB}\equiv\text{tr}\sqrt{\rho_{AB}}(X_i\otimes Y_j)$ and $\gamma_{kl}^{CD}\equiv\text{tr}\sqrt{\rho_{CD}}(Z_k\otimes W_l)$. Let $\Gamma_{AB}=(\gamma_{ij}^{AB}),\Gamma_{CD}=(\gamma_{kl}^{CD})$, which may be regarded as some kind of correlation matrices for the state $\rho_{AB}$ and $\rho_{CD}$, respectively. Then we have
\begin{equation}\label{4.3}
\begin{aligned}
\sqrt{\rho_{AB}\otimes\rho_{CD}}= & \sum_{i=0}^{m^{2}-1}\sum_{j=0}^{n^{2}-1}\sum_{k=0}^{u^{2}-1}\sum_{l=0}^{v^{2}-1}\gamma_{ij}^{AB}\gamma_{kl}^{CD}X_{i}\otimes Y_{j}\otimes Z_{k}\otimes W_{l}, \\
\sqrt{\rho_{BC,AD}}= & \sum_{j=0}^{n^{2}-1}\sum_{k=0}^{u^{2}-1}\sum_{i=0}^{m^{2}-1}\sum_{l=0}^{v^{2}-1}\gamma_{ij}^{AB}\gamma_{kl}^{CD}Y_{j}\otimes Z_{k}\otimes X_{i}\otimes W_{l},
\end{aligned}
\end{equation}
where the matrix $\Gamma_{BC,AD}=(\gamma_{jk,il}^{BC,AD})=(\gamma_{ij}^{AB}\gamma_{kl}^{CD})=\Gamma_{AB}^{t}\otimes\Gamma_{CD}$.

\begin{theorem}\label{theorem2}
For $\rho_{AB}$ and $\rho_{CD}$ represented as Eq. \eqref{4.2}, we have
\begin{equation}\label{4.4}
\begin{aligned}
N_H^{b}(\rho_{AB}\otimes\rho_{CD})=1-\min_{G}\mathrm{tr}G\Gamma_{BC,AD}\Gamma_{BC,AD}^{t}G^{t}\leq1-\sum_{s=1}^{nu}\lambda_{s},
\end{aligned}
\end{equation}
where $G\equiv(g_{h(jk)})$ is an $nu\times n^{2}u^{2}$ dimensional matrix with $g_{h(jk)}\equiv\mathrm{tr}\Pi_{h}^{BC}Y_{j}\otimes Z_{k}(h=0,1,\ldots,nu-1;(jk)=ju^{2}+k,j=0,1,\ldots,n^{2}-1,k=0,1,\ldots,u^{2}-1)$, $\Gamma_{BC,AD}=(\gamma_{ij}^{AB}\gamma_{kl}^{CD})_{jk,il}=\Gamma_{AB}^{t}\otimes\Gamma_{CD}$ is an $n^{2}u^{2}\times m^{2}v^{2}$ dimensional matrix, and $\{\lambda_{s}:s=1,2,\ldots,n^{2}u^{2}\}$ are the eigenvalues of the matrix $\Gamma_{BC,AD}\Gamma_{BC,AD}^{t}$ listed in increasing order.

Furthermore, without loss of generality, if $\rho_{B}$ is nondegenerate with spectral projections $\{|e_{B}\rangle\langle e_{B}|\}$, then $\Pi^{BC}=\{\Pi_{e}^{B}\otimes \Pi_{f}^{C}\}=\{|e_{B}\rangle\langle e_{B}|\otimes \Pi_{f}^{C}\}$ and
\begin{equation}\label{4.5}
\begin{aligned}
N_H^{b}(\rho_{AB}\otimes\rho_{CD})&=1-\mathrm{tr}B\Gamma_{AB}^{t}\Gamma_{AB}B^t\times\min_{C}\mathrm{tr}C\Gamma_{CD}\Gamma_{CD}^{t}C^{t}\\
&\leq1-\mathrm{tr}B\Gamma_{AB}^{t}\Gamma_{AB}B^t\times(\sum_{s^{'}=1}^{u}\lambda^{'}_{s^{'}}),
\end{aligned}
\end{equation}
where $B\equiv(b_{ej})$ is an $n\times n^{2}$ dimensional matrix with $b_{ej}\equiv\mathrm{tr}|e_{B}\rangle \langle e_{B}|Y_{j}=\langle e_{B}|Y_{j}|e_{B}\rangle$, $C\equiv(c_{fk})$ is a $u\times u^{2}$ dimensional matrix with $c_{fk}\equiv\mathrm{tr}\Pi^{C}_{f}Z_{k}$, and $\{\lambda^{'}_{s^{'}}:s^{'}=1,2,\ldots,u^2\}$ are the eigenvalues of the matrix $T_{CD}T_{CD}^{t}$ listed in increasing order.

In particular, if $u=2$, then

\begin{equation}\label{4.6}
\begin{aligned}
&N_{H}^{b}(\rho_{AB}\otimes\rho_{CD})=1-\mathrm{tr}B\Gamma_{AB}^{t}\Gamma_{AB}B^t\times(\|\mathbf{r}_{CD}\|^{2}+r_{\min}),
\end{aligned}
\end{equation}
where $\mathbf{r}_{CD}=(\gamma_{00}^{CD},\gamma_{01}^{CD},\ldots,\gamma_{0(v^2-1)}^{CD}),\|\mathbf{r}_{CD}\|^{2}=\sum_l(\gamma_{0l}^{CD})^2,$ and $r_{\min}$ is the smallest eigenvalue of the $(3\times3)$-dimensional matrix $RR^t$ with $R=(\gamma_{kl})_{k=1,2,3;l=0,1,\ldots,v^2-1}$.

Finally, if $\rho_{B}$ and $\rho_{C}$ are both nondegenerate with spectral projections $\{|e_{B}\rangle\langle e_{B}|\}$ and $\{|f_{C}\rangle\langle f_{C}|\}$ respectively, then
\begin{equation}\label{4.7}
\begin{aligned}
N_H^{b}(\rho_{AB}\otimes\rho_{CD})=1-\mathrm{tr}B\Gamma_{AB}^{t}\Gamma_{AB}B^t\times\mathrm{tr}C\Gamma_{CD}\Gamma_{CD}^{t}C^{t},
\end{aligned}
\end{equation}
where $C\equiv(c_{fk})$ is a $u\times u^{2}$ dimensional matrix with $c_{fk}\equiv\mathrm{tr}|f_{C}\rangle\langle f_{C}|Z_{k}=\langle f_{C}|Z_{k}|f_{C}\rangle$.
\end{theorem}
\begin{proof*}
Firstly, we have
\begin{equation}\label{4.8}
\begin{aligned}
N_{H}^{b}(\rho_{AB}\otimes\rho_{CD})=N_H(\rho_{BC,AD}).
\end{aligned}
\end{equation}
From Eq. \eqref {4.3}, we obtain
\begin{equation*}
\begin{aligned}
\Pi^{BC}(\sqrt{\rho_{BC,AD}})=&\sum_h\sum_{ijkl}\gamma_{ij}^{AB}\gamma_{kl}^{CD}\Pi_h^{BC}(Y_j\otimes Z_k)\Pi_h^{BC}\otimes X_i\otimes W_l\\
=&\sum_h\sum_{ijkl}\gamma_{ij}^{AB}\gamma_{kl}^{CD}g_{h(jk)}\Pi_h^{BC}\otimes X_i\otimes W_l\\
=&\sum_h\sum_{ijj^{'}kk^{'}l}\gamma_{ij}^{AB}\gamma_{kl}^{CD}g_{h(jk)}g_{h(j^{'}k^{'})}Y_{j^{'}}\otimes Z_{k^{'}}\otimes X_i\otimes W_l.
\end{aligned}
\end{equation*}

Consequently,
\begin{equation*}
\begin{aligned}
&\mathrm{tr}\sqrt{\rho_{BC,AD}}\Pi^{BC}(\sqrt{\rho_{BC,AD}})\\
=&\sum_h\sum_{ijj^{'}kk^{'}l}\gamma_{ij}^{AB}\gamma_{kl}^{CD}g_{h(jk)}g_{h(j^{'}k^{'})}\gamma_{ij^{'}}^{AB}\gamma_{k^{'}l}^{CD}\\
=&\sum_h\sum_{ijj^{'}kk^{'}l}g_{h(jk)}\gamma_{jk,il}^{BC,AD}\gamma_{j^{'}k^{'},il}^{BC,AD}g_{h(j^{'}k^{'})}\\
=&G\Gamma_{BC,AD}\Gamma_{BC,AD}^{t}G^{t}.
\end{aligned}
\end{equation*}

Noting that $\mathrm{tr}\sqrt{\rho_{BC,AD}}\Pi^{BC}(\sqrt{\rho_{BC,AD}})=\mathrm{tr}[\Pi^{BC}(\sqrt{\rho_{BC,AD}})]^2$ and Eq. \eqref{4.8}, we have
\begin{equation*}
\begin{aligned}
N_{H}^{b}(\rho_{AB}\otimes\rho_{CD})=&\max_{\Pi^{BC}}\|\sqrt{\rho_{AB}\otimes\rho_{CD}}-\Pi^{BC}(\sqrt{\rho_{AB}\otimes \rho_{CD}})\|^2\\
=&1-\min_G G\Gamma_{BC,AD}\Gamma_{BC,AD}^{t}G^{t},
\end{aligned}
\end{equation*}

Because of $GG^t=I^B\otimes I^C$, we have $\min_G G\Gamma_{BC,AD}\Gamma_{BC,AD}^{t}G^{t}\geq\sum_{s=1}^{nu}\lambda_{s}$, the desired inequality \eqref{4.4} follows.

Furthermore, without loss of generality, if $\rho_{B}$ is nondegenerate, then

\begin{equation*}
\begin{aligned}
&N_{H}^b(\rho_{AB}\otimes\rho_{CD})\\
=&\max_{\Pi^{C}}\|\sqrt{\rho_{AB}\otimes\rho_{CD}}-(\Pi^{B}\otimes\Pi^{C})(\sqrt{\rho_{AB}\otimes\rho_{CD}})\|^2\\
=&\max_{\Pi^{C}}\|\sqrt{\rho_{AB}}\otimes\sqrt{\rho_{CD}}-\Pi^{B}(\sqrt{\rho_{AB}})\otimes\Pi^{C}(\sqrt{\rho_{CD}})\|^2\\
=&1-\min_{\Pi^{C}}\mathrm{tr}(\sqrt{\rho_{AB}}\Pi^{B}(\sqrt{\rho_{AB}})\otimes\sqrt{\rho_{CD}}\Pi^{C}(\sqrt{\rho_{CD}}))\\
=&1-\mathrm{tr}\sqrt{\rho_{AB}}\Pi^{B}(\sqrt{\rho_{AB}})\times\min_{\Pi^{C}}\mathrm{tr}\sqrt{\rho_{CD}}\Pi^{C}(\sqrt{\rho_{CD}})\\
=&1-\mathrm{tr}B\Gamma_{AB}^{t}\Gamma_{AB}B^t\times\min_{C}\mathrm{tr}C\Gamma_{CD}\Gamma_{CD}^{t}C^{t}\\
\leq&1-\mathrm{tr}B\Gamma_{AB}^{t}\Gamma_{AB}B^t\times(\sum_{s^{'}=1}^{u}\lambda^{'}_{s^{'}}).
\end{aligned}
\end{equation*}

If $u=2$, the identity $\sum_{f=0}^1\Pi_{f}^{C}=I^{C}$ implies that $c_{0k}=-c_{1k}$ for $k=1,2,3$. If we denote $\mathbf{c}\equiv\sqrt{2}(c_{01},c_{02},c_{03})$, then from $\sum_{k=0}^3a_{0k}^2=1$ and $c_{00}=c_{10}=1/\sqrt{2}$, we have $\|\mathbf{c}\|=1$. Conversely, for any $\mathbf{c}\equiv\sqrt{2}(c_{01},c_{02},c_{03})$ with $\|\mathbf{c}\|=1$, the operator
$$\frac{1}{\sqrt{2}}\frac{I^C}{\sqrt{2}}+\sum_{k=1}^3c_{fk}\frac{\sigma_k}{\sqrt{2}}$$
is a pure state. Now, $$C=(c_{fk})=\frac{1}{\sqrt{2}}\left(
                                                       \begin{array}{cc}
                                                         1 & \mathbf{c} \\
                                                         1 & -\mathbf{c} \\
                                                       \end{array}
                                                     \right),
$$
and
$$\Gamma_{CD}=(\gamma_{kl}^{CD})=\left(
           \begin{array}{c}
             \mathbf{r}_{CD} \\
             R \\
           \end{array}
         \right),
$$
with $\mathbf{r}_{CD}=(\gamma_{00}^{CD},\gamma_{01}^{CD},\ldots,\gamma_{0(v^2-1)}^{CD})$, which is a $v^2$-dimensional row vector, and $R=(\gamma_{kl})_{k=1,2,3;l=0,1,\ldots,v^2-1}$, which is a $(3\times v^2)$-dimensional matrix; then we obtain $$\mathrm{tr}C\Gamma_{CD}\Gamma_{CD}^{t}C^{t}=\|\mathbf{r}_{CD}\|^{2}+\mathbf{c}RR^t\mathbf{c}^t.$$

This implies the desired result, Eq. \eqref{4.6}.

Finally, if $\rho_{B}$ and $\rho_{C}$ are both nondegenerate with spectral projections $\{|e_{B}\rangle\langle e_{B}|\}$ and $\{|f_{C}\rangle\langle f_{C}|\}$ respectively, then the Eq. \eqref{4.7} is obtained obviously.
\qed
\end{proof*}
\section{Examples}\label{sec:4}
In this section, we calculate the measure-induced nonbilocality for the given pure state and mixed state respectively.
\begin{example}
According to Theorem \ref{theorem1}, for any Bell states, e.g. $|\Phi_{AB}\rangle\otimes|\Phi_{CD}\rangle=\frac{1}{\sqrt{2}}(|00\rangle+|11\rangle)_{AB}\otimes\frac{1}{\sqrt{2}}(|00\rangle+|11\rangle)_{CD}$, we have
$$N_{H}^{b}(|\Phi_{AB}\rangle\langle\Phi_{AB}|\otimes|\Phi_{CD}\rangle\langle\Phi_{CD}|)=1-4\times(\frac{1}{\sqrt{2}})^{4}\times(\frac{1}{\sqrt{2}})^{4}=\frac{3}{4}.$$
\end{example}
\begin{example}
In contrast, the measure-induced nonbilocality may not vanishes when input states are both separable. For the classical state $\rho^{c}=\frac{1}{2}|0\rangle\langle0|\otimes|0\rangle\langle0|+\frac{1}{2}|1\rangle\langle1|\otimes|1\rangle\langle1|$, we have $N_{H}^b(\rho^{c}_{AB}\otimes\rho^{c}_{CD})=\frac{3}{4}$. Since
$$\sqrt{\rho^{c}}=\frac{1}{\sqrt{2}}|0\rangle\langle0|\otimes|0\rangle\langle0|+\frac{1}{\sqrt{2}}|1\rangle\langle1|\otimes|1\rangle\langle1|,$$ and
$$\sqrt{\rho^{c}}=\frac{1}{\sqrt{2}}\frac{I}{\sqrt{2}}\otimes\frac{I}{\sqrt{2}}+\frac{1}{\sqrt{2}}\frac{\sigma_{3}}{\sqrt{2}}\otimes\frac{\sigma_{3}}{\sqrt{2}},$$
where $$\Gamma_{AB}=\Gamma_{CD}=\left(
    \begin{array}{cccc}
      \frac{1}{\sqrt{2}} & 0 & 0 & 0 \\
      0 & 0 & 0 & 0 \\
      0 & 0 & 0 & 0 \\
      0 & 0 & 0 & \frac{1}{\sqrt{2}} \\
    \end{array}
  \right),$$ then we have
\begin{equation*}
\begin{aligned}
\Gamma_{BC,AD}=\Gamma_{AB}^{t}\otimes\Gamma_{CD}=diag(\frac{1}{2},O_{2},\frac{1}{2},O_{8},\frac{1}{2},O_{2},\frac{1}{2}).
\end{aligned}
\end{equation*}
where $O_{n}$ represents the $n\times n$ zero matrix.

According to Theorem \ref{theorem2}, if we choose one of the most optimal von Neumann measurements $$\Pi^{BC}=\{H^{\otimes 2}|00\rangle\langle00|H^{\otimes 2},H^{\otimes 2}|01\rangle\langle01|H^{\otimes 2},H^{\otimes 2}|10\rangle\langle10|H^{\otimes 2},H^{\otimes 2}|11\rangle\langle11|H^{\otimes 2}\},$$ where $H$ denote the Hadamard matrix $$H=\frac{1}{\sqrt{2}}\left(
                               \begin{array}{cc}
                                 1 & 1 \\
                                 1 & -1 \\
                               \end{array}
                             \right),
$$
then we can obtain the matrix $$G=\frac{1}{2}\left(
\begin{array}{cccccccccccccccc}
1 & 1 & 0 & 0 & 1 & 1 & 0 & 0 & 0 & 0 & 0 & 0 & 0 & 0 & 0 & 0\\
1 & -1 & 0 & 0 & 1 & -1 & 0 & 0 & 0 & 0 & 0 & 0 & 0 & 0 & 0 & 0\\
1 & 1 & 0 & 0 & -1 & -1 & 0 & 0 & 0 & 0 & 0 & 0 & 0 & 0 & 0 & 0\\
1 & -1 & 0 & 0 & -1 & 1 & 0 & 0 & 0 & 0 & 0 & 0 & 0 & 0 & 0 & 0\\
\end{array}
\right),
$$
and
$$G\Gamma_{BC,AD}\Gamma_{BC,AD}^{t}G^{t}=\frac{1}{16}\left(
                                                      \begin{array}{cccc}
                                                        1 & 1 & 1 & 1 \\
                                                        1 & 1 & 1 & 1 \\
                                                        1 & 1 & 1 & 1 \\
                                                        1 & 1 & 1 & 1 \\
                                                      \end{array}
                                                    \right)
.$$

Finally, we have
\begin{align*}
N_{H}^b(\rho^{c}_{AB}\otimes\rho^{c}_{CD})=1-\min_G\mathrm{tr}G\Gamma_{BC,AD}\Gamma_{BC,AD}^{t}G^{t}=\frac{3}{4}.
\end{align*}
\end{example}
\begin{example}
For the Bell diagonal state $$\beta^{AB}=\frac{1}{3}(|\Psi^+\rangle\langle\Psi^+|+|\Psi^-\rangle\langle\Psi^-|+|\Phi^+\rangle\langle\Phi^+|),$$ where $|\Psi^\pm\rangle=(|00\rangle\pm|11\rangle)/\sqrt{2},|\Phi^\pm\rangle=(|01\rangle\pm|10\rangle)/\sqrt{2}$; then by the spectral theorem,$$\sqrt{\beta^{AB}}=\frac{1}{\sqrt{3}}(|\Psi^+\rangle\langle\Psi^+|+|\Psi^-\rangle\langle\Psi^-|+|\Phi^+\rangle\langle\Phi^+|).$$

According to the results of Example 1 in \cite{42}, we have
$$\min_{A}A\Gamma_{AB}\Gamma_{AB}^tA^t=\frac{1}{4}(h^2+\min_j{d_j^2})=\frac{1}{4}((\sqrt{3})^2+(\frac{1}{\sqrt{3}})^2)=\frac{5}{6}.$$
So $$N_{H}(\beta^{AB})=1-\min_{A}A\Gamma_{AB}\Gamma_{AB}^tA^t=\frac{1}{6},$$
and $$1-\min_{\Pi^{A}}(\mathrm{tr}\sqrt{\beta_{AB}}\Pi^{A}(\sqrt{\beta_{AB}}))^{2}=1-(\frac{5}{6})^2=\frac{11}{36}.$$

Now we calculate the value of $N_{H}^b(\beta^{BA}\otimes\beta^{AB})$. Since
$$\sqrt{\beta^{AB}}=\frac{1}{\sqrt{3}}(|\Psi^+\rangle\langle\Psi^+|+|\Psi^-\rangle\langle\Psi^-|+|\Phi^+\rangle\langle\Phi^+|)
=\frac{1}{\sqrt{3}}\left(
                     \begin{array}{cccc}
                       1 & 0 & 0 & 0 \\
                       0 & \frac{1}{2} & \frac{1}{2} & 0 \\
                       0 & \frac{1}{2} & \frac{1}{2} & 0 \\
                       0 & 0 & 0 & 1 \\
                     \end{array}
                   \right)
,$$ and
$$\sqrt{\beta^{AB}}=\frac{\sqrt{3}}{2}\frac{I}{\sqrt{2}}\otimes\frac{I}{\sqrt{2}}+\frac{1}{2\sqrt{3}}\frac{\sigma_{1}}{\sqrt{2}}\otimes\frac{\sigma_{1}}{\sqrt{2}}+\frac{1}{2\sqrt{3}}\frac{\sigma_{2}}{\sqrt{2}}\otimes\frac{\sigma_{2}}{\sqrt{2}}+\frac{1}{2\sqrt{3}}\frac{\sigma_{3}}{\sqrt{2}}\otimes\frac{\sigma_{3}}{\sqrt{2}},$$
where $$\Gamma_{AB}=\left(
    \begin{array}{cccc}
      \frac{\sqrt{3}}{2} & 0 & 0 & 0 \\
      0 & \frac{1}{2\sqrt{3}} & 0 & 0 \\
      0 & 0 & \frac{1}{2\sqrt{3}} & 0 \\
      0 & 0 & 0 & \frac{1}{2\sqrt{3}} \\
    \end{array}
  \right),$$ then we have
\begin{equation*}
\begin{aligned}
\Gamma_{AA,BB}&=\Gamma_{BA}^t\otimes\Gamma_{AB}\\
&=diag(\frac{3}{4},\frac{1}{4},\frac{1}{4},\frac{1}{4},\frac{1}{4},\frac{1}{12},\frac{1}{12},\frac{1}{12},\frac{1}{4},\frac{1}{12},\frac{1}{12},\frac{1}{12},\frac{1}{4},\frac{1}{12},\frac{1}{12},\frac{1}{12}).
\end{aligned}
\end{equation*}

According to Theorem \ref{theorem2}, if we choose the von Neumann measurement $$\Pi^{BC}=\{|\Psi^+\rangle\langle\Psi^+|,|\Psi^-\rangle\langle\Psi^-|,|\Phi^+\rangle\langle\Phi^+|,|\Phi^-\rangle\langle\Phi^-|\},$$
then we can obtain the matrix $$G=\frac{1}{2}\left(
\begin{array}{cccccccccccccccc}
1 & 0 & 0 & 0 & 0 &  1 & 0 & 0 & 0 & 0 & -1 & 0 & 0 & 0 & 0 & 1\\
1 & 0 & 0 & 0 & 0 & -1 & 0 & 0 & 0 & 0 &  1 & 0 & 0 & 0 & 0 & 1\\
1 & 0 & 0 & 0 & 0 &  1 & 0 & 0 & 0 & 0 &  1 & 0 & 0 & 0 & 0 & -1\\
1 & 0 & 0 & 0 & 0 & -1 & 0 & 0 & 0 & 0 & -1 & 0 & 0 & 0 & 0 & -1\\
\end{array}
\right),
$$
and
$$G\Gamma_{AA,BB}\Gamma_{AA,BB}^{t}G^{t}=\frac{1}{144}\left(
                                                      \begin{array}{cccc}
                                                        21 & 20 & 20 & 20 \\
                                                        20 & 21 & 20 & 20 \\
                                                        20 & 20 & 21 & 20 \\
                                                        20 & 20 & 20 & 21 \\
                                                      \end{array}
                                                    \right)
.$$

Finally, we have
\begin{align*}
N_{H}^b(\beta_{BA}\otimes\beta_{AB})=1-\min_G\mathrm{tr}G\Gamma_{AA,BB}\Gamma_{AA,BB}^{t}G^{t}\geq1-\frac{7}{12}=\frac{5}{12}.
\end{align*}
\begin{equation*}
\begin{aligned}
N_{H}^{b}(\beta_{BA}\otimes\beta_{AB})=&\max_{\Pi^{AA}}\|\sqrt{\beta_{BA}\otimes\rho_{AB}}-\Pi^{AA}(\sqrt{\beta_{BA}\otimes\beta_{AB}})\|^{2}\geq\frac{5}{12}\\
\geq&\max_{\Pi^{A}}\|\sqrt{\beta_{BA}\otimes\beta_{AB}}-(\Pi^{A}\otimes\Pi^{A})(\sqrt{\beta_{BA}\otimes\beta_{AB}})\|^2\\
=&1-\min_{\Pi^{A}}(\mathrm{tr}\sqrt{\beta_{AB}}\Pi^{A}(\sqrt{\beta_{AB}}))^{2}=\frac{11}{36}\\
\geq&1-\min_{\Pi^{A}}~\mathrm{tr}\sqrt{\beta_{AB}}\Pi^{A}(\sqrt{\beta_{AB}})\\
=&N_{H}(\beta_{AB})=\frac{1}{6}.
\end{aligned}
\end{equation*}
\end{example}
\section{Conclusion}\label{sec:5}
Here, we define a quantifier of the nonbilocal correlation and
discuss its computational properties.
We obtain a strong connection between the Measurement-Induced Nonlocality and the measurement-induced nonbilocal correlation,
which implies our measure can detect more nonlocal correlations.
We have obtained the exact analytical formulas of the nonbilocality measure when inputs are pure
states. However, for the mixed-input case, we get a upper bound
of the quantifier.
\acknowledgments
Thanks for comments. The work is supported by
National Science Foundation of China under Grant No. 11771011.

\end{document}